\documentclass[preprint]{elsarticle}

\usepackage[colorlinks=true,urlcolor=blue,linkcolor=blue,citecolor=blue]{hyperref}
\usepackage{chemformula} 
\usepackage{natbib}
\usepackage{graphicx}
\usepackage{float}
\usepackage{caption}
\usepackage{booktabs}
\usepackage{soul}
\usepackage{amssymb}
\usepackage{subfigure}

\journal{Journal of \LaTeX\ Templates}

\biboptions{numbers,sort&compress}

\begin{document}

\begin{frontmatter}

\title{Differences in \ch{Sb2Te3} growth by pulsed laser and sputter deposition}


\author[1]{Jing Ning\corref{correspondingauthor}}
\cortext[correspondingauthor]{Corresponding author}
\ead{jing\_ ning@mymail.sutd.edu.sg}

\author[1]{J. C. Martinez}
\author[2]{Jamo Momand}
\author[2]{Heng Zhang}
\author[3]{Subodh C. Tiwari}
\author[4]{Fuyuki Shimojo}
\author[3]{Aiichiro Nakano} 
\author[3]{Rajiv K. Kalia} 
\author[3]{Priya Vashishta} 
\author[5]{Paulo S. Branicio}
\author[2]{Bart J. Kooi}
\author[1]{Robert E. Simpson\corref{correspondingauthor}}
\ead{robert\_ simpson@sutd.edu.sg}

\address[1]{Singapore University of Technology and Design (SUTD), 8 Somapah Road, 487372, Singapore}
\address[2]{Zernike Institute for Advanced Materials, University of Groningen (UG), Nijenborgh 4, Groningen 9747 AG, The Netherlands}
\address[3]{Collaboratory for Advanced Computing and Simulation, University of Southern California (USC), Los Angeles, CA-90089, U. S. A.}
\address[4]{Department of Physics, Kumamoto University, Kumamoto 860-8555, Japan}
\address[5]{Mork Family Department of Chemical Engineering and Materials Science, University of Southern California, Los Angeles, CA-90089, U. S. A.}

\begin{abstract}
High quality van der Waals chalcogenides are important for phase change data storage, thermoelectrics, and spintronics. 
Using a combination of statistical design of experiments and density functional theory, we clarify how the out-of-equilibrium van der Waals epitaxial deposition methods can improve the crystal quality of \ch{Sb2Te3} films. 
We compare films grown by radio frequency sputtering and pulsed laser deposition (PLD).
The growth factors that influence the crystal quality for each method are different.
For PLD grown films a thin  amorphous \ch{Sb2Te3} seed layer most significantly influences the crystal quality. 
In contrast, the crystalline quality of films grown by sputtering is rather sensitive to the deposition temperature and less affected by the presence of a seed layer.
This difference is somewhat surprising as both methods  are out-of-thermal-equilibrium plasma-based methods. 
Non-adiabatic quantum molecular dynamics simulations show that this difference originates from the density of excited atoms in the plasma.
The PLD plasma is more intense and with higher energy than that used in sputtering, and this increases the electronic temperature of the deposited atoms, which concomitantly increases the adatom diffusion lengths in PLD.
In contrast, the adatom diffusivity is dominated by the thermal temperature for sputter grown films.  
These results explain the wide range of \ch{Sb2Te3} and superlattice crystal qualities observed in the literature.
These results indicate that, contrary to popular belief, plasma-based deposition methods are suitable for growing high quality crystalline chalcogenides.
\end{abstract}


\end{frontmatter}

\section{Introduction}
\ch{Sb2Te3}, \ch{Bi2Te3}, and \ch{Bi2Se3} are typical van der Waals (vdW) layered chalcogenide crystals that can be used in phase storage memory devices, thermoelectrics, and spintronics\citep{wuttig2007phase, lencer2008map, siegrist2011disorder, zhang2009topological, chen2011molecular, osterhage2014thermoelectric, zheng2013enhanced}. 
\ch{Sb2Te3} exists at one extreme of the \ch{Sb2Te3-GeTe} pseudobinary tie line, along which many well-known  phase change materials (PCMs) exist that exhibit significant contrast between amorphous and crystalline states, low switching energy, and read-write repeatability\citep{wuttig2007phase, lencer2008map, siegrist2011disorder}. 
They are also important topological insulators due to a stable Dirac cone at the gamma point in the electronic band structure\citep{zhang2009topological, chen2011molecular}. 
These materials are also thermoelectrics with high thermoelectric figures of merit\citep{osterhage2014thermoelectric, zheng2013enhanced}. 
With so many technological applications, it is important to understand how to grow high quality \ch{Sb2Te3} and the factors that may used to optimise its growth. 

\begin{figure}
\centering
\includegraphics[width=\textwidth]{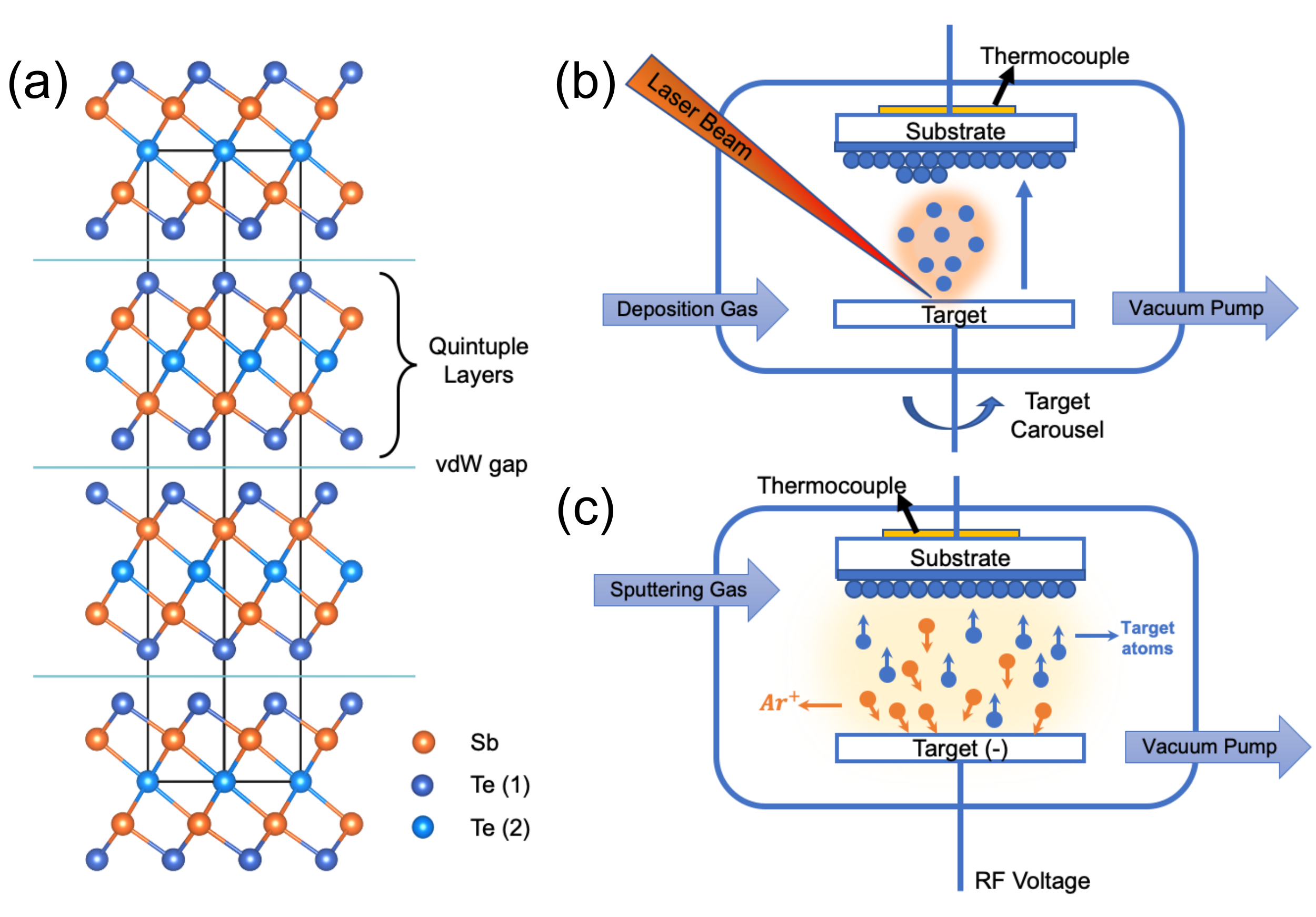}
\caption{(a) R$\bar{3}m$ crystal structure of \ch{Sb2Te3} within a hexagonal unit cell. 
Schematics of 
(b) pulsed laser and 
(c) sputter deposition methods.
}
\label{fig_1}
\end{figure}

The \ch{Sb2Te3} crystal has a  R$\bar{3}m$ crystal structure with a unit cell consisting of three stacked quintuple layers (QLs) --Te$_1$--Sb--Te$_2
$--Sb--Te$_1$--vdW--  along its c-axis\citep{anderson1974refinement}. 
The QLs are bonded by a weak vdW interaction, whilst inter-layer bonding is strong because of the dominant covalent character within each block\citep{mishra1997electronic}. 
The crystal structure is shown in Figure~\ref{fig_1}(a). 

The exotic properties of \ch{Sb2Te3} depend on its crystal structure, thus growing high-quality single crystals or well-textured crystalline thin films is essential\citep{wuttig2014exploring}. 
In PCMs, \ch{GeTe-Sb2Te3} superlattices switch at substantially lower energies than alloys of the same composition.
This increase in switching efficiency is due atomic motions  being limited to the superlattice layer interfaces, which reduces entropic losses\cite{simpson2011interfacial}. 
\ch{Sb2Te3} is often used as a seed layer for vdW epitaxy of phase change superlattices\citep{simpson2011interfacial, saito2016two}, while also being useful for applying biaxial strain to enhance the performance of phase change memory devices\citep{zhou2016phase}. 
\citeauthor{zhou2016phase} and \citeauthor{kalikka2016evolutionary} used the \ch{GeTe-Sb2Te3} lattice mismatch  to further improve the performance of these phase change superlattices. 
Biaxial stress influences diffusive atomic motions and the activation energy for Ge atoms to switch into the vdW interface\citep{zhou2016phase,kalikka2016evolutionary}. 
Clearly, for phase change memory applications it is important to grow \ch{Sb2Te3} layers with high quality interfaces with the GeTe. 

The thermoelectric properties of \ch{Bi2Te3}, \ch{Sb2Te3}, and their superlattices have the highest known ZT values in the research literature. 
The performance of a thermoelectric device is assessed by a figure of merit, the ZT value, where Z is the scale of a material's thermal properties and T is the absolute temperature.
\citeauthor{venkatasubramanian2001thin} reported a remarkable ZT$\sim$2.4 in  a \ch{Bi2Te3-Sb2Te3} superlattice device. 
Accurately growing ultra-short-period superlattice structures allowed the phonon and electron transport to be controlled and significantly enhance the ZT value\citep{venkatasubramanian2001thin}. 
Similarly, a high degree of ordering is effective at improving the thermal power.
\citeauthor{tan2012fabrication} fabricated and integrated a highly oriented \ch{Sb2Te3} thin film with a layered structure into a thermoelectric power generation device, which exhibited much better performance than randomly oriented or $(0\ 1\ 5)$  oriented \ch{Sb2Te3} films\citep{tan2012fabrication, shen2017enhancing,tan2013enhanced}. 

These R$\bar{3}m$ chalcogenide crystals are also known as topological insulators.  
Moreover, the same superlattice structures that are applied to phase change memories are also Dirac semimetals\citep{tominaga2014ferroelectric}.
It is, therefore, possible to form topological insulator--normal insulator memories or topological switching devices\citep{sa2012topological}.
However these exotic properties strongly depend on the crystal structure and the number of atomic layers (thickness) of the structure. 
Hence, the structure must be fabricated with exquisite control.
Highly textured \ch{Sb2Te3} thin films were grown to investigate the electrical transport properties as a function of thickness and temperature;  atomically flat single crystalline \ch{Sb2Te3} exhibited the topologically insulating behaviour\citep{zhang2014investigation, wang2010atomically}.
Thus high quality crystal structures allowed the electronic properties to be engineered with topological states\citep{saha2016pulsed, neupane2014observation}. 

Crystalline thin film vdW solids can be prepared by a variety of methods\citep{greene1982ion}, such as mechanical exfoliation\citep{shahil2012micro}, chemical vapour deposition (CVD)\citep{novoselov20162d}, and physical vapour deposition (PVD)\citep{behera2018sb2te3, boschker2017textured, hilmi2016epitaxial}. 
Exfoliation can obtain perfect single crystal layers but it is a very laborious and time-consuming process and can only cover small areas\citep{geim2013van}; chemical vapour deposition (CVD) requires precursors and elevated temperatures, which causes impurities and mechanical instabilities due to thermal expansion and stress\citep{novoselov20162d}. 
PVD methods are preferred because they can be highly automated, are industrially compatible, and allow \emph{in situ} observation of the crystal growth.
Recently, a number of publications have reported PVD growth of \ch{Sb2Te3}. 
Magnetron sputtering is commonly used to deposit phase change superlattice materials\citep{behera2018sb2te3, saito2016two, kowalczyk2018impact, saito2020high}, and \ch{Sb2Te3}-based alloys have been realised in industrial fabrication lines. 
However, it is hard to control the uniformity and accurate atomistic stacking of crystals. 
Molecular beam epitaxy (MBE) provides a greater degree of control of the deposited atoms. 
However, the deposition rate is slow and typically between 0.3 and 0.5 nm/min\citep{boschker2017textured, momand2015interface}, and it cannot be used to deposit directly from alloy sources. 
An alternative and somewhat straightforward growth method is pulsed laser deposition (PLD)\citep{hilmi2016epitaxial, hilmi2017research, vermeulen2018strain}. 
A laser is used to ablate the surface of a target, forming a plume of plasma, from which the material is deposited on a substrate.
It is capable of providing stoichiometric deposition from alloy targets with high deposition rates. 
Indeed, PLD has been used to successfully grow a wide gamut of materials, including Transition Metal Dichalcogenides (TMDCs) \citep{hilmi2017research, hilmi2016epitaxial, vermeulen2018strain, serrao2015highly, loh2015one, ullah2016pulsed, eason2007pulsed}.

Anecdotally, the chalcogenide research community seems to accept that there may be differences in crystal quality for these different deposition methods, but an accurate comparison of their similarities and the factors that influence the crystal growth is currently lacking.
Moreover, understanding these dependencies might help to explain why different research groups seem to grow different crystals despite their growth temperatures being similar. 
For example, there are some reports that the sputter grown \ch{Sb2Te3} is actually Te deficient\citep{kowalczyk2018impact, saito2020high}, whilst other groups do not see this effect, and there are even reports that high quality \ch{Sb2Te3} can be grown at  temperatures as low as 140 $^{\circ}$C  while other reports show that temperatures as high as 300 $^{\circ}$C are necessary\citep{hilmi2017research,zhou2016phase}.
\citeauthor{boschker2017growth} discussed the differences between sputtering, PLD and MBE for \ch{Ge2Sb2Te5} fabrication, but the origin of these differences is unclear\citep{boschker2017growth}. 
Here, we quantify the differences in the growth conditions and provide insight into the origin of the conflicting observations in the literature.  

It is clear that \ch{Sb2Te3} crystal films are important but it remains unclear whether the deposition factors that influence the crystal quality are the same for different growth methods. 
Since a large number of factors are involved in the growth process, it is useful to employ statistical designs to find the factors that have a significant impact on the crystal quality. 
One-factor-at-a-time (OFAT) method is typically applied to study the effect of one experimental variable, but this is inefficient, time-consuming, and laborious, especially when there are more than 4 factors. 
In contrast, factorial design is a systematic method to determine the relationship between factors affecting a process and the output of that process. 
It has been applied in fields as diverse as industrial product design, materials development,  chemical and medical research\citep{otto2001product, ilzarbe2008practical}. 
If one is willing to decrease the sensitivity to interactions of multiple factors, then the number of measurements can be reduced to a half, quarter, or one-eighth by exploiting fractional factorial design (FFD) methodology. 

In this paper, we present a systematic and quantitative study on the growth of highly $(0\ 0\ l)$ oriented \ch{Sb2Te3} with vdW epitaxy. 
Herein, we use factorial designs and analysis of variance (ANOVA) to compare the factors that influence the crystal quality of PLD and sputter grown films. 
Non-adiabatic quantum molecular dynamics (NAQMD) simulations are also performed to study the effect of the PLD plasma on the atomic diffusivity of \ch{Sb2Te3} at the temperatures typically used to grow \ch{Sb2Te3} vdW growth experiments.  
We also used ground state density functional molecular dynamics (DF/MD) to illustrate and compare how ionised Sb and Te atoms crystallise with and without a seed layer.
Our findings shine new light on the vdW epitaxy growth dependence on deposition methods, thus opening a way to grow high quality \ch{Sb2Te3} vdWs crystals, its superlattices, and related materials such as TMDCs.

\section{Methods and Experiments}
\subsection{Design of Experiment \& Analysis}

Fractional factorial design and ANOVA was used to quantify the significance of growth factors on the \ch{Sb2Te3} crystal quality. 
For PLD, the factors considered were temperature, deposition pressure, laser fluence, distance between target and substrate, and the seed layer. 
The aim was to optimise the crystal quality of oriented \ch{Sb2Te3} crystalline films. 
These experimental factors and settings are shown in Table~\ref{tbl_1} and each combination is assigned a label from 1 to 16. 
The results are referenced against  a film prepared with factors at an intermediate level with values half-way between the high and low settings. 
This reference sample is labelled R in Table~\ref{tbl_1}. 
Factors are symbolized from A to E, each factor has two levels, where $+1$ is high level, and the $-1$ is low. 
For sputtered films, it was only necessary to study the seed layer and temperature effects because these were previously found to be the two most significant factors influencing the \ch{Sb2Te3} crystal quality and the degree of out-of-plane orientation\citep{behera2018sb2te3}. 
These were calculated from XRD patterns using the objective function shown in Eq.~\ref{eq_1}, 
which is able to optimise both the crystal quality and the growth rate.
Thus, to maximise the crystal quality and orientation, we must maximize $ Q_{(0\ 0\ l)}$.

\begin{equation}
	Q_{(0\ 0\ l)}=\frac{1}{N}\sum_{i=1}^{N} \frac{I_{(0\ 0\ l)}}{FWHM_{(0\ 0\ l)}}
  \label{eq_1}
\end{equation}

In  Eq.~\ref{eq_1}, $I_{(0\ 0\ l)}$ is the intensity of $(0\ 0\ l)$ peaks, $FWHM_{(0\ 0\ l)}$ is the full width at half maximum of the corresponding peaks. 
Peak (0 0 3), (0 0 6), (0 0 9), (0 0 15), (0 0 18) were used to quantify the crystal quality. 
The intensity and the corresponding FWHM of the diffraction peaks were analysed after pseudo-Voigt fitting with the MDI Jade software.
ANOVA was applied to screen for the most significant growth parameters.

\begin{table}\LARGE
\caption{Experimental matrix of deposition variables in PLD}
\label{tbl_1}
\resizebox{\textwidth}{!}{
\begin{tabular}{cccccc}
\toprule
 & \textbf{A (Temperature/ $^{\circ}C$)} & \textbf{B (Pressure/ $Pa$)}  & \textbf{C (Laser fluence/ $J\cdot\ cm^{-2}$)} & \textbf{D (Target-Substrate distance/ $cm$)}&\textbf{E=ABCD (Seed layer)} \\
\midrule\Large
$1$ & $240\ (+1)$ &$ 50\ (+1)$ & $0.9\ (+1)$& $7\ (+1)$& $3\ nm \ (+1)$\\
$2$ & $240\ (+1)$ &$ 50\ (+1)$ & $0.9\ (+1)$& $6\ (-1)$& $No\ (-1)$\\
$3$ & $240\ (+1)$ &$ 50\ (+1)$ & $0.6\ (-1)$& $7\ (+1)$& $No\ (-1)$\\
$4$ & $240\ (+1)$ &$ 50\ (+1)$ & $0.6\ (-1)$& $6\ (-1)$& $3\ nm (+1)$\\
$5$ & $240\ (+1)$ &$ 5\ (-1)$ & $0.9\ (+1)$& $7\ (+1)$& $No (-1)$\\
$6$ & $240\ (+1)$ &$ 5\ (-1)$ & $0.9\ (+1)$& $6\ (-1)$& $3\ nm\ (+1)$\\
$7$ & $240\ (+1)$ &$ 5\ (-1)$ & $0.6\ (-1)$& $7\ (+1)$& $3\ nm\ (+1)$\\
$8$ & $240\ (+1)$ &$ 5\ (-1)$ & $0.6\ (-1)$& $6\ (-1)$& $No\ (-1)$\\
$9$ & $180\ (-1)$ &$ 50\ (+1)$ & $0.9\ (+1)$& $7\ (+1)$& $No\ (-1)$\\
$10$ & $180\ (-1)$ & $50\ (+1)$ & $0.9\ (+1)$& $6\ (-1)$& $3\ nm\ (+1)$\\
$11$ & $180\ (-1)$ & $50\ (+1)$ & $0.6\ (-1)$& $7\ (+1)$& $3\ nm\ (+1)$\\
$12$ & $180\ (-1)$ & $50\ (+1)$ & $0.6\ (-1)$& $6\ (-1)$& $No\ (-1)$\\
$13$ & $180\ (-1)$ & $5\ (-1)$ & $0.9\ (+1)$& $7\ (+1)$& $3\ nm\ (+1)$\\
$14$ & $180\ (-1)$ & $5\ (-1)$ & $0.9\ (+1)$& $6\ (-1)$& $No\ (-1)$\\
$15$ & $180\ (-1)$ & $5\ (-1)$ & $0.6\ (-1)$& $7\ (+1)$& $No\ (-1)$\\
$16$ & $180\ (-1)$ & $5\ (-1)$ & $0.6\ (-1)$& $6\ (-1)$& $3\ nm\ (+1)$\\
Reference & $210\ (0)$ & $12\ (0)$ & $0.78\ (0)$& $6\ (-1)$& $3\ nm\ (+1)$\\
\bottomrule
\end{tabular}}
\end{table}	

\subsection{Growth of \ch{Sb2Te3} thin films}


\ch{Sb2Te3} thin films were grown on 20~nm thick silicon nitride grids using a PLD system from TSST with the growth conditions, shown in Table~\ref{tbl_1}. 
The alloy target was bought from KTECH. 
The purity was 99.999\%. 
It was ablated by a KrF excimer laser beam ($\lambda =248\ $nm). 
The base pressure of the vacuum chamber was better than $5\times{10}^{-6}\ $Pa. 
The deposition temperature, argon gas pressure, laser fluence, target-substrate distance and seed layer were set to the combinations, shown in Table~\ref{tbl_1}. 
The \ch{Sb2Te3} seed layer was formed by applying 200 laser pulses to the target at room temperature to produce a $3.5 \pm{0.5}\ $nm film. 
The substrate was subsequently heated at $5\ ^{\circ}$C/min to the growth temperature. 
A further 1800 pulses were applied at a laser repetition frequency of 1 Hz. The 2000 pulses were directly carried out if there was no seed layer.
A schematic depicting the PLD method is shown in Figure~\ref{fig_1}(b).

\ch{Sb2Te3} films were also prepared on silicon $(1\ 0\ 0)$ substrates by RF magnetron sputtering (AJA International). The chamber base pressure was kept better than $6\ \times{10}^{-5}$ Pa and the samples were prepared in an Ar pressure of 0.63 Pa. 
The deposition power was 8\ W. The native oxide of Si was removed from the Si substrate by argon plasma etching for one hour. 
Then a 3 nm-thick \ch{Sb2Te3} seed layer was grown in amorphous state at room temperature and annealing was performed before the the thin films were grown at high temperature. 
After deposition the films were cooled naturally to room temperature overnight in the vacuum chamber to avoid oxidation. 
A schematic showing the key features of sputter deposition is shown in Figure~\ref{fig_1} (c). 

\subsection{Characterization}
The surface morphology and grain size were measured by Scanning Electron Microscopy (SEM, FEI Nova NanoSEM 650) and Transmission Electron Microscopy (TEM, JEOL 2010). 
The SEM measurements were performed at a 20 kV potential, whilst 200 kV was used for TEM. 
A transmission detector in the SEM was used to collect bright field scanning transmission electron microscope (STEM) images.
The films were grown directly on a 20~nm thick silicon nitride membrane grids. 
This enabled inspection of the plan view of the sample surface using both the STEM mode in SEM and the TEM. 
For the TEM cross-sectional images, Focused Ion Beam (FIB, Helios G4 CX DualBeam\textsuperscript{TM}) milling was necessary. 
The elemental composition of \ch{Sb2Te3} films was analysed by Energy Dispersive X-ray spectroscopy (EDX) and the out of plane vdW layers were imaged and studied by selected area electron diffraction (SAED). 
The crystal orientation and grain size were analysed in the SEM using electron backscatter diffraction (EBSD) in transmission-mode, which provides a higher spatial resolution than reflection-mode\citep{keller2012transmission, van2016optimum}. 
OIM Analysis\textsuperscript{TM} was used to analyse the Kikuchi pattern collected by the t-EBSD scan. 
Additionally, the crystal structure of the samples was measured using X-ray diffraction (XRD, Bruker D8 Discover diffractometer) with Cu$K_\alpha$($\lambda =1.5418$ \AA) radiation in a symmetric Bragg-Brentano geometry for $\theta$--$2\theta$ ranging from 5$^{\circ}$ to 60$^{\circ}$, which is sensitive to $(0~0~l)$ orientation.

\subsection{Modelling \& Simulaitons}

DF/MD calculations were performed using the Vienna $ab\  initio$ Simulation Package (VASP)\citep{Kresse1993} with projected augmented wave (PAW) pseudopotentials\citep{Kresse1999} and Perdew-Burke-Ernzerhof generalised gradient approximation (PBE--GGA) exchange-correlation functional\citep{Perdew1996}. 
The spin-orbit interaction was neglected and the vdW interaction correction method of Grimme was included\citep{Brooks2006}.
The \ch{Sb2Te3} structure geometry relaxation was carried out at 0~K using the \ch{Sb2Te3} unit cell-- see Figure \ref{fig_1}(a).
We used a plane-wave basis with a 240~eV cutoff. 
For MD simulations, an \ch{Sb2Te3} supercell ($3\times 3\times1$)  consisting of 54~Sb and 81~Te atoms was used.
A canonical ensemble (NVT) was used to update the atomic positions every 3 fs.
The energy of the ensemble was computed at the $\Gamma$ point in the Brillouin zone (k=0). 
The plane-wave cutoff energy was 175 eV. 
The temperature of the model was controlled by velocity rescaling.  
An NVT ensemble was used to heat the relaxed crystal structure to 600 K at a rate of 100 K/ps. 
An NPT ensemble was then used to equilibrate  the cell at 600~K.
Subsequently, the NPT ensemble was used to heat the structure from 600 K to 1500 K at a rate of  150 K/ps, and then to hold at 1500 K for 10~ps for melting.
To create the amorphous state the system was quenched from 1500~K to 800~K  in 7.78~ps.
To demonstrate the effect of a seed layer on the crystal growth process, an atomic plane of Te atoms was fixed in the centre of the simulation.
In the PLD process, the deposited atoms are ionised in the plasma plume.
Considering the long computation time of DF/MD crystallisation simulations, we exaggerate the effect of electronic excitation on crystal growth by using an artificially proportion of excited electrons by removing the 10\% highest energy valence band electrons. 
This corresponds to states in the energy window from 1 eV below the valence band maximum (VBM) to the VBM, which is mainly the Te p-orbitals\citep{martinez2020origin}. 
A similar approach has been used by other to study excited state dynamics in \ch{Ge2Sb2Te5}\citep{li2011role}

Crystallinity was evaluated using Bond Order Parameter (BOP) proposed by \citeauthor{steinhardt1983bond}, which is commonly used to identify the different local atomic crystal structures and to study structural phase transitions\citep{steinhardt1983bond, wang2005melting, kalikka2012nucleus}. 
The local structure around atom $i$ is defined by a set of spherical harmonics, where $Y_{lm}(\widehat{\textbf{r}}_{ij})$, $\widehat{\textbf{r}}_{ij}$ is the vector between atom $i$ and one of its neighbours $j$.
$N_b$ is the total number of neighbours around the $i^{th}$ atom. 
The global bond order parameter, $Q_l$, can be then calculated by averaging according to:  
\begin{equation}
	Q_{l} \equiv {\sqrt{ {\frac{4\pi}{2l+1} \sum_{m=-1}^{l}
	 {\left| 
	 {\frac{1}{N_b}
	 {\sum_{j=1}^{N(i)}
	 Y_{lm}(\widehat{\textbf{r}}_{ij})}}\right| }^2}}}
	\label{eq_3}
\end{equation}
where $l=4$ for a local atomic cubic structure. 

The DF/MD model artificially approximates the excited state by removing the most energetic electrons from the crystal. 
A more careful crystallisation simulation involving excited state dynamics is challenging because existing time-dependent DFT codes are much more computationally intensive than ground state codes, like VASP.  
However, we also ran much shorter (too short for crystallisation) molecular dynamic simulations using the QXMD software \citep{Shimojo2019}. 
For non-adiabatic quantum molecular (NAQMD) simulations, we utilized GGA\citep{Perdew1996} with non-linear core correction\citep{Louie1982} for the exchange correlation energy and vdW correction based on DFT-D scheme\citep{Brooks2006}. 
We have also employed the PAW method\citep{Blochl1994} and generate functions for the 5s, 5p, and 4d states of Sb and Te. 
We have used a cutoff energy of 30 and 250 Ry 
for the electronic pseudo-wave functions and the pseudo-charge density, respectively. 
Again, the $\Gamma$ point is used in the Brillouin zone to compute the energies. 
Interatomic forces are computed quantum mechanically based on the Hellmann-Feynman theorem. 
The equations of motion are integrated numerically with a time step of 50 a.u.(1.21 fs). 
The NAQMD method describes electronic excitations in the framework of linear-response time-dependent DFT\citep{Casida1995a, Tully1990a}. 
In addition, non-adiabatic transitions between excited electronic states assisted by molecular motions are treated with a surface hopping approach\citep{Duncan2007}. 
Due to the use of excited-state forces, photoexcitation also modifies ground-state electronic structures. 
Details on the QXMD software are described elsewhere\citep{Shimojo2019, Shimojo2013}. 
The NAQMD simulations were performed with a 180 atoms supercell by performing a $4\times3\times1$ duplication of a crystalline orthorhombic cell of \ch{Sb2Te3}. 
\ch{Sb2Te3} was converted to an orthorhombic unit cell by taking a projection along the $[2\ 1\ 0]$ direction.
Initial excitation was performed by removing electrons from HOMO to HOMO$-n$ bands and placing them in LUMO to LUMO$+n$ bands, where $n$ is the number of excited bands.
This NAQMD simulation was carried out with variable proportions (n = 2.6, 5.2, 10.3 and 15\%) valence electrons in excited states, which correspond to carrier densities of 0.45, 0.91, 1.82 and 2.44 ($\times 10^{22}\ cm^{-3}$), respectively.
The temperature of the simulation was kept well below the melting temperature at 500~K.

\section{Results}
 \ch{Sb2Te3} thin films were grown on silicon nitride TEM window grids (substrates) from a \ch{Sb2Te3} alloy target by PLD. 
 The composition measurements on different regions of different samples showed an  average composition in at.$\%$ of $39.6 \pm{1.6}$ Sb and $60.4 \pm{2.0}$ Te, which indicates good stoichiometric transfer.
 This implies that the composition is relatively insensitive to the PLD  growth temperature between 180~$^{\circ}$C to 240~$^{\circ}$C. 

\begin{figure}[!htb]
\centering
\includegraphics[width=\textwidth]{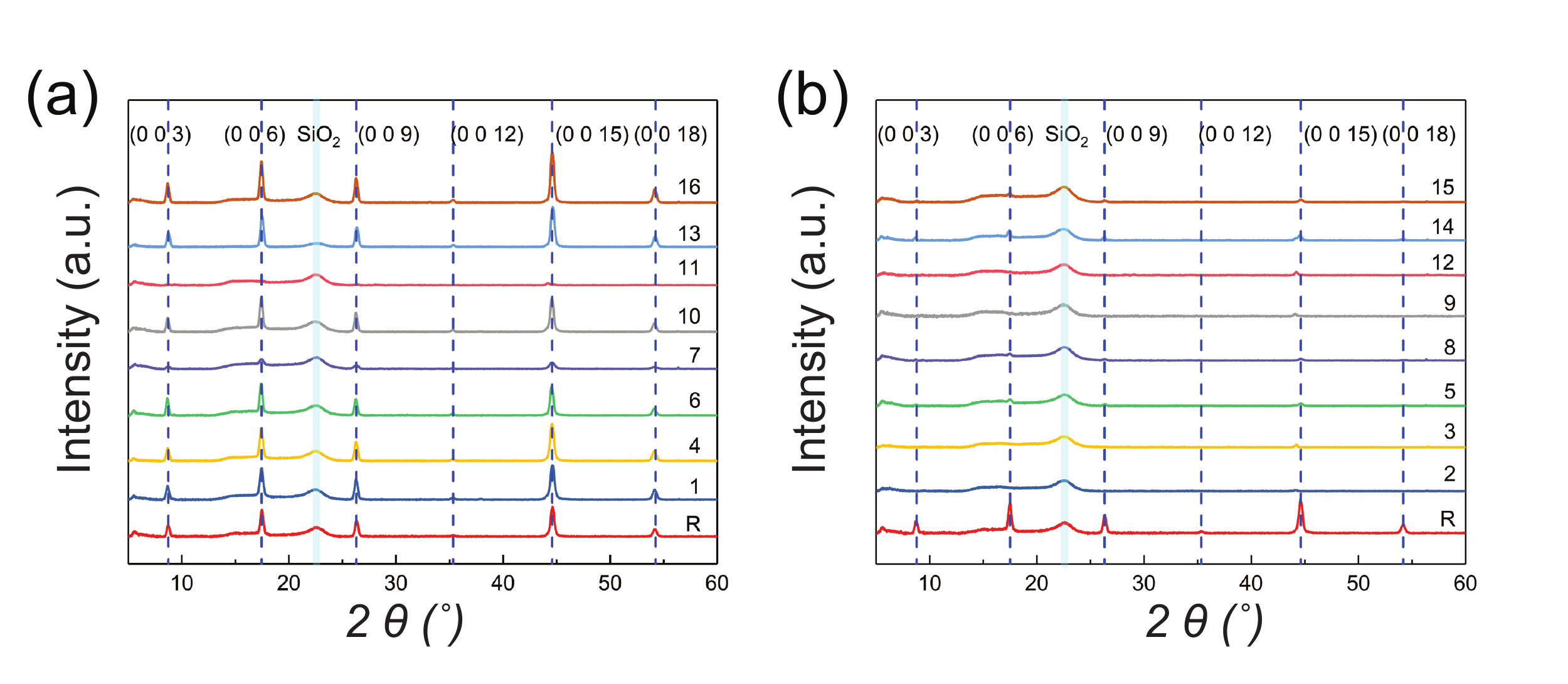}
\caption{XRD patterns of \ch{Sb2Te3} films deposited by PLD with the different growth conditions used in the FFD study (see Table~\ref{tbl_1}). 
The films were divided into two groups: those with a seed layer (a), and those without a seed layer (b). 
The blue-dashed lines indicate the expected $(0\ 0\ l)$ diffraction peak angles for \ch{Sb2Te3}. 
The cyan line shows the angle at which X-ray scattering from the \ch{SiO2} glass slide is maximum.}
\label{fig_2}
\end{figure}

Qualitatively, we found that depositing a 3~nm thick \ch{Sb2Te3} seed layer increases the propensity for the PLD grown films to have a $(0\ 0\ l)$ crystallographic orientation.
To test this we split the 16 XRD diffraction patterns that were used in the DoE study (Table~\ref{tbl_1}) into two groups: those with a seed layer and those without a seed layer. 
These patterns are presented in Figure~\ref{fig_2}(a) (with an \ch{Sb2Te3} seed layer) and  \ref{fig_2}(b) (no seed layer).
Layered c-axis \ch{Sb2Te3} crystals were only observed when the seed layer was deposited at room temperature prior to growing the thicker layer at a higher temperature.
This is clear from Figure~\ref{fig_2}(a), which only shows $(0~0~l)$ peaks. 
This $(0\ 0\ l)$ orientation is especially strong for growth conditions \#13 and \#16, where sharp and intense peaks are observed. 
Electron microscopy analysis also showed that films \#13  and \#16 were particularly flat and clearly exhibited hexagonally-shaped grains; again indicating a high degree of $(0\ 0\ l)$ texture. 
In contrast, the group of samples that were grown without a seed layer exhibited broad and low intensity diffraction peaks, as is clear  in Figure~\ref{fig_2}(b). 
Note, samples \#2, \#3, \#9 and \#12 were noncontinuous films with isolated islands, while sample \#8 and \#15 were continuous but exhibited  lower and broader diffraction peaks.
Also see Figure S1 in the Supporting Information. 

\begin{figure}[!htb]
\centering
\includegraphics[width=\textwidth]{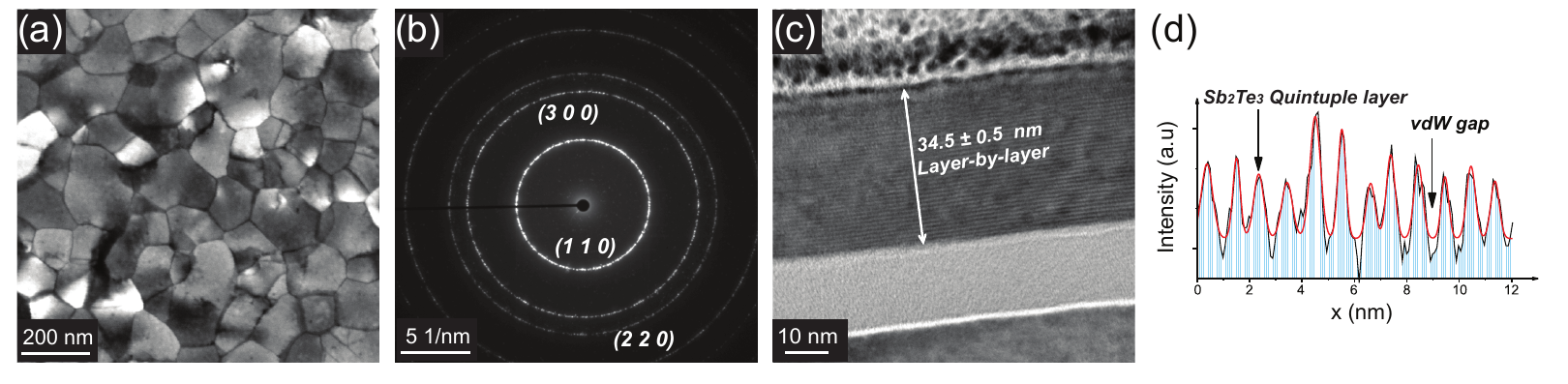}
\caption{Morphology and structure of the reference PLD-grown \ch{Sb2Te3} sample. 
(a) Clear domains and grain boundaries are seen in the bright-field STEM image plan view. 
(b) Bright and uniform rings corresponding to the $(h~k~0)$ planes are seen in the SAED pattern. 
(c) Cross-sectional view of the \ch{Sb2Te3} film showing the vdW layers are observed. 
(d) The quintuple layers and vdW gaps are more clearly seen in a line-intensity plot scanned in the out-of-plane direction. 
The growth condition for the reference sample is listed in Table~\ref{tbl_1}.}
\label{fig_3}
\end{figure}

\begin{figure}[!htb]
\centering
\includegraphics[width=\textwidth]{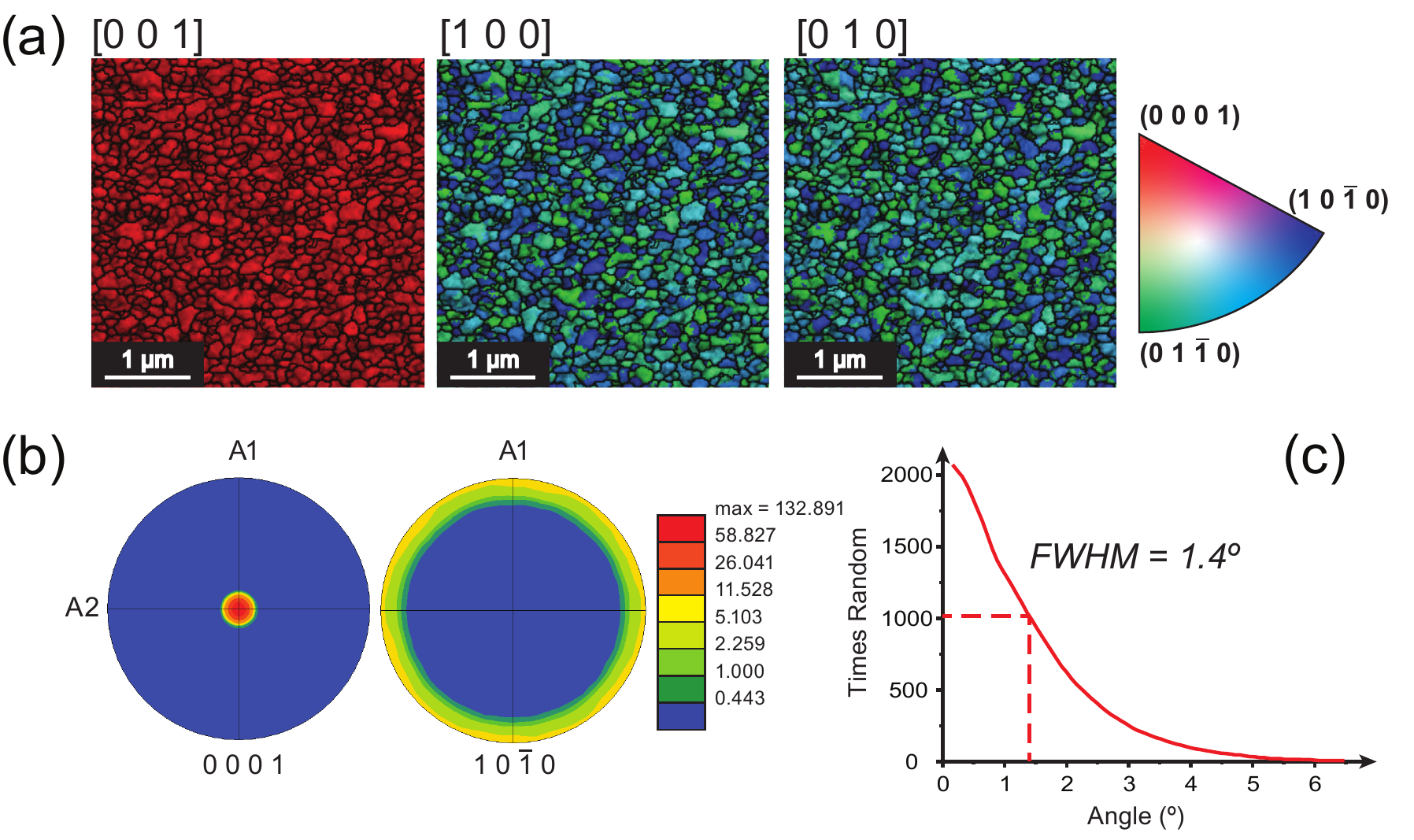}
\caption{(a) Transmission-EBSD crystallographic orientation maps of the reference PLD-grown \ch{Sb2Te3} film. 
The texture maps for the [0 0 1], [1 0 0], and [0 1 0] directions.
(b)Pole figures for $(0\ 0\ 0\ 1)$ and $(1\ 0\ \bar{1}\ 0\ )$ crystallographic planes.
(c) Misorientation distribution for the out-of-plane (0\ 0\ 0\ 1) crystal orientation confirms a highly textured \ch{Sb2Te3} thin film.}
\label{fig_4}
\end{figure}

To quantify the improvement in $(0~0~l)$ orientation and crystal quality, it is important to establish a reference sample against which the other samples can be compared. 
For this reference we used pressure, temperature, and fluence values midway between the extremes values used in the FFD study. 
We decided to use a 3~nm thick \ch{Sb2Te3} seed layer and a 6~cm target-substrate distance to ensure that a continuous film was deposited.
The reference film was grown on a silicon nitride membrane at room temperature. 
Subsequently we deposited \ch{Sb2Te3} at a rate of 1.0 nm/min or 0.016 nm/pulse at $210^{\circ}$C. 
The resultant \ch{Sb2Te3} film thickness was 34.5$\pm$0.5~nm.
The growth condition is shown in Table~\ref{tbl_1} and the corresponding diffraction pattern is labelled R in Figure~\ref{fig_2}. 
The reference smaple $(0\ 0\ l)$ XRD peaks are clearly seen  and the hexagonal morphology of the grains and degree of orientation can be seen by SEM, TEM and t-EBSD in Figure~\ref{fig_3} and Figure~\ref{fig_4}, respectively.
The STEM image also shows a relatively flat film with relatively large domains and clear grain boundaries (Figure~\ref{fig_3}(a)). 
The SAED pattern shows rings corresponding to $(h~k~0)$ planes, which are parallel with electron beam direction (Figure~\ref{fig_3}(b)). 
This indicates that the reference film has a layered crystal structure with random orientation in-plane and $(0\ 0\ l)$ texture out-of-plane. 
Additionally, the in-plane lattice constant, which was calculated from the position of these diffraction rings, is 4.266 \AA. 
This  is consistent with published measurements for bulk \ch{Sb2Te3}, which is 4.264~\AA \citep{anderson1974refinement}.
The layered structure of the reference sample is shown clearly in the cross-sectional TEM image (Figure~\ref{fig_3}(c)).
The vdW gaps are clearly observed and are spaced 1~nm apart, as shown in the line profile (Figure~\ref{fig_3}(d)).
This is in excellent agreement with the expected  spacing between $(0~0~l)$ layers of a highly oriented \ch{Sb2Te3} crystal structure. 
Thus the cross-sectional and SAED images both confirm the out-of-plane texture. 
The inverse pole figures of [0 0 1], [1 0 0] and [0 1 0] directions also show that the crystal grains are well-textured along the c-axis with random in-plane orientation.
The EBSD crystal orientation maps for this reference film are shown in Figure~\ref{fig_4}(a). 
The centre of the $(0~0~0~1)$ crystallographic plane and the uniform annulus around the edge of the $(1~0~\bar{1}~0)$ pole figure is a signature of strong out-of-plane and random in-plane orientation, as shown in Figure~\ref{fig_4}(b). 
This agrees with the XRD and SAED measurements shown in Figure~\ref{fig_2} and \ref{fig_3}(b). 
The crystal grains have a narrow distribution around the $(0~0~0~1)$  plane, the misorientation distribution full width at half maximum (FWHM) is just 1.4$^{\circ}$, as seen in Figure~\ref{fig_4}(c).

Note, the SAED images for all of the samples listed in Table~\ref{tbl_1} are shown in Supplementary Figure S2.
Darker and more blurred diffraction rings indicate poorer crystal quality.

To quantify the importance of the seed layer  relative to other growth factors, its statistical significance needs to be measured. 
We used ANOVA to assess the significance of the deposition factors on $ Q_{(0\ 0\ l)}$, which was calculated using Eq.~\ref{eq_1} from the XRD patterns shown in Figure~\ref{fig_2} and compared against the reference sample. 
We found that the seed layer is the most and only significant factor in PLD \ch{Sb2Te3} crystal growth. 
This is interesting because \citeauthor{behera2018sb2te3} reported that the buffer type (W \& Mo buffer layer), temperature, power and pressure significantly affect the quality of sputtered \ch{Sb2Te3}. 
Moreover, in \citeauthor{behera2018sb2te3}'s study the influence of an amorphous \ch{Sb2Te3} layer was not tested.
Therefore, to directly  compare  sputtering and PLD, we further performed a temperature and \ch{Sb2Te3} seed layer ANOVA test for sputtered \ch{Sb2Te3} films.

\begin{figure}[!htb]
\centering
\includegraphics[width=\textwidth]{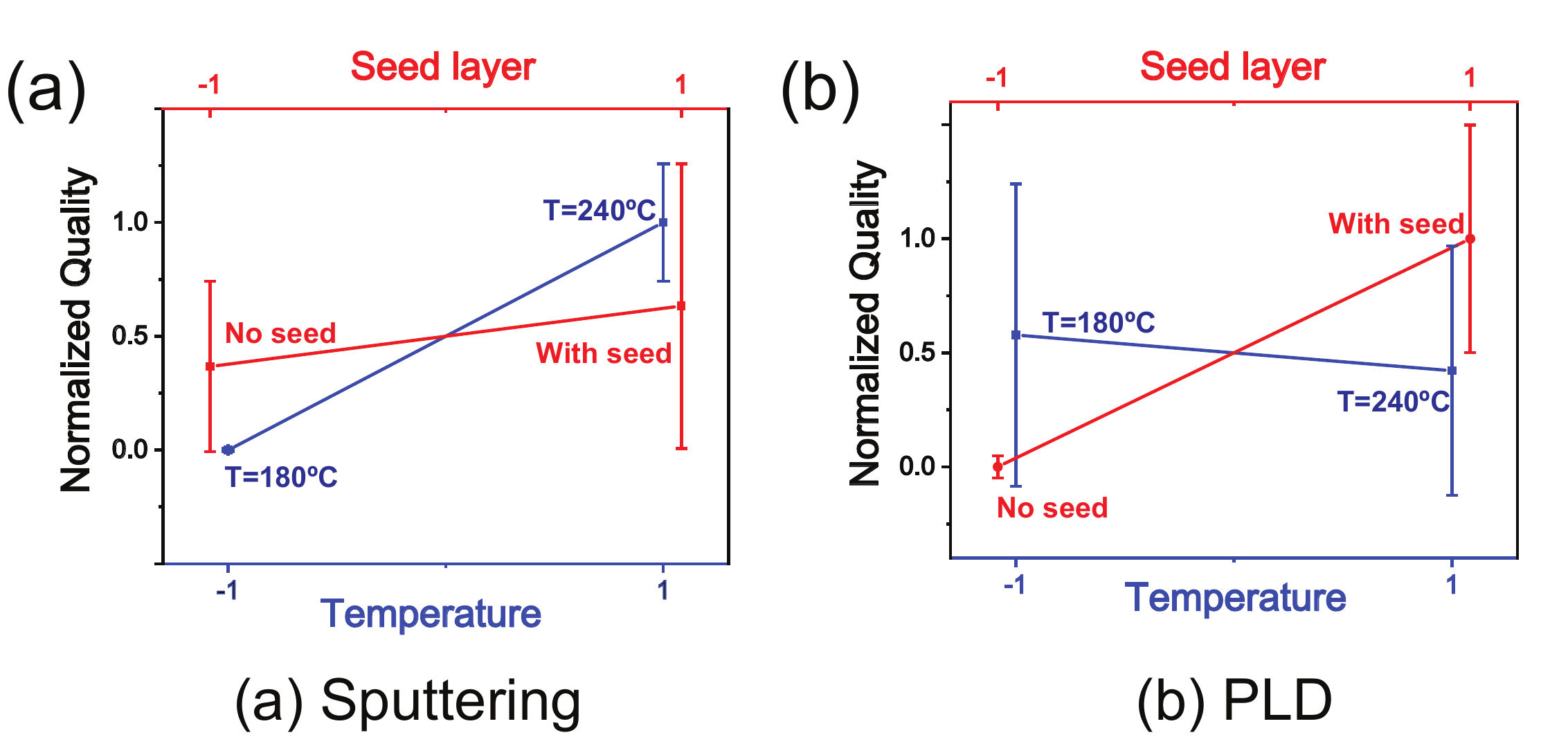}
\caption{Main Effect plots for seed layer and temperature effects in (a) sputtering and (b) PLD. 
ANOVA was used to determine the  influence of temperature (blue squares) and seed layer (red circles) on the crystal quality. 
}
\label{fig_5}
\end{figure}

For sputter-grown \ch{Sb2Te3}, we found that temperature significantly affects the \ch{Sb2Te3} crystal quality and although the seed layer does have an effect, its statistical significance is lower.
This is shown in  Figure~\ref{fig_5}(a), which shows a large quality difference between the two temperature levels and small deviation in crystal-quality for samples prepared at the same temperature (small within-group variation). 
This dependence on temperature is in striking contrast to PLD, where the seed layer has a far more significant effect on the PLD-grown crystal quality, as shown in Figure~\ref{fig_5}(b). 
I.e. the red curve shows that the crystal quality shows a large positive improvement when a seed layer is included, but the effect of temperature (blue) is insignificant.
Moreover, across the temperature range studied, a higher crystal quality of \ch{Sb2Te3} is observed at higher temperatures for sputter grown films, but the quality decreases with temperature for the PLD-grown films. 
We will go on to show that this difference is caused by electronic excitation of the atoms in PLD. 


\section{Discussion}

The amorphous \ch{Sb2Te3} seed layer facilitates highly-oriented $(0\ 0\  l)$ texture in both PLD and sputtering, as shown in Figure~\ref{fig_2} and ~\ref{fig_5}. 
\citeauthor{saito2016two} discussed the advantages of a seed layer on the sputter depositing previously. 
We can now see that an amorphous \ch{Sb2Te3} buffer is even more important for PLD grown films.
\citeauthor{saito2016two} argued that the Te layer terminates the dangling bonds on the substrate surface, which prevents good layered heteroepitaxial growth on it\citep{koma1992van, koma1999van}. 
The passivation effect of the Te-atomic layer was also studied in \citeauthor{hippert2020growth}'s work. 
They found that depositing a thin Te buffer layer allowed high quality and out-of-plane oriented \ch{Sb2Te3} films to be grown on a range of substrates, including amorphous silicon, silicon oxide, \ch{TiN} and \ch{WSi}\citep{hippert2020growth}. 
Hydrogen and Sb were also used to passivate dangling bonds and performed vdW epitaxial growth of \ch{GeTe-Sb2Te3} superlattices on Si(1 1 1) substrates by MBE\citep{momand2016atomic}. 
Growing the seed layer at room temperature and then annealing at high temperatures prior to growing thicker \ch{Sb2Te3} provides a highly-oriented template for subsequent epitaxial growth.\citep{saito2016two, saito2015self}. 
This seed layer may also reduce the lattice mismatch with the substrate such that a quasi-homoepitaxial growth is possible. 

\begin{figure}[!htb]
\centering
\includegraphics[width=\textwidth]{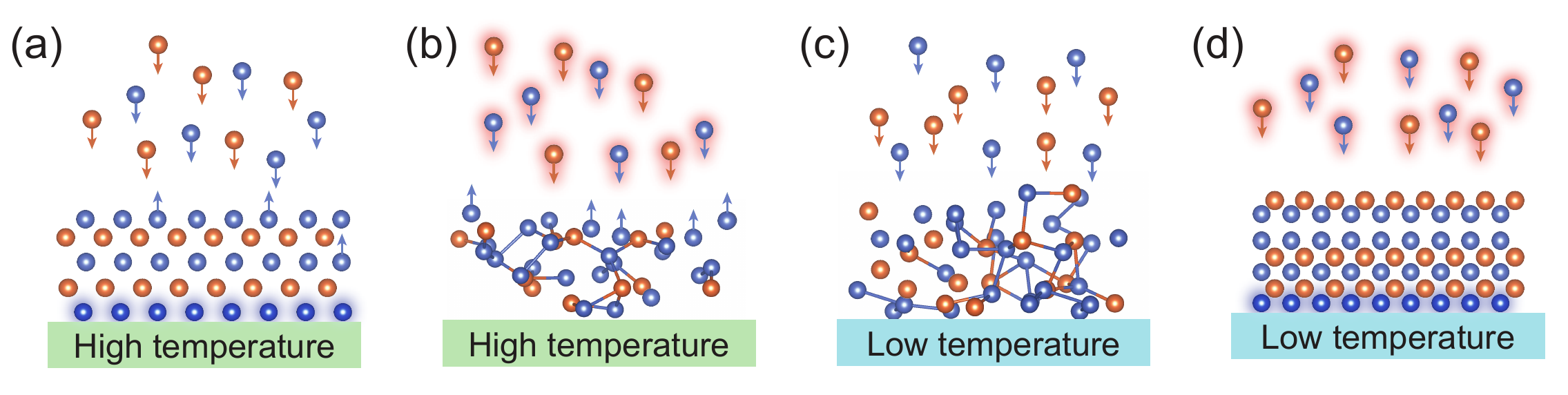}
\caption{Growth models for different growth conditions. 
(a) depicts deposition on a seed layer at high temperature. 
This is commonly used in sputter deposition of \ch{Sb2Te3}\citep{saito2015self, saito2016two, behera2018sb2te3, saito2020high}
(b) shows no seed layer, electronic excited Sb and Te atoms directly deposited at high temperature. 
This has been demonstrated by our PLD growth.
(c) shows no seed layer at low temperatures.
This is the usual case for room temperature sputtering of amorphous \ch{Sb2Te3}\citep{fang2013effects}
(d) depicts deposition of Sb and Te atoms in an electronic excited state (high electronic temperature) at a low thermal temperature. 
This is possible for PLD-growth of \ch{Sb2Te3}.
In all figures, Orange and blue balls correspond to Sb and Te atoms respectively.
The red glow means the atoms are in an excited state.}
\label{fig_6}
\end{figure}

Figure~\ref{fig_5} clearly shows that temperature  plays a different role in PLD and sputtering. 
In equilibrium deposition methods, the adatom diffusivity is controlled by the substrate temperature. 
The surface diffusion coefficient $D_s$ is defined by
\begin{equation}
	D_s=\upsilon{a^2}\exp\left(-\frac{E_a}{k_BT}\right)
\label{eq_2}
\end{equation}
where $\upsilon$ is the jump frequency, $a$ the jump distance, $E_a$ the activation energy, $k_B$ the Boltzmann constant and $T$ the temperature. 
It is clear from Eq.~\ref{eq_2} that increasing the deposition temperature increases the surface diffusivity, and this is important because the adatom must diffuse long distances to find a low energy and stable positions in the crystal structure\citep{eason2007pulsed}. 
We assumed that the sputtered films at high temperatures will form layered crystal structures, especially for films with \ch{Sb2Te3} seed layers, as shown in Figure~\ref{fig_6}(a).
However, the Te sticking coefficient is also related to the temperature and it is impractical to use temperature to control adatom diffusivity.

Tellurium has a high vapour pressure at modertately high temperatures.
For example, the sticking coefficient is substantially reduced above 600~K \citep{mzerd1995crystal}, thus films become Te deficient. 
Therefore 600~K is the highest practical growth temperature for \ch{Sb2Te3} using quasi-equilibrium deposition methods, such as MBE. 
This low sticking coefficient may also explain some of the non-stoichiometric films reported recently in the literature\citep{kowalczyk2018impact, saito2020high} and the lower growth rates at temperatures above 533~K\citep{hilmi2017research}. 
Figure~\ref{fig_6}(b) depicts the case where the substrate temperature is too high, and there is no seed layer. 
The deposited film is likely to be polycrystalline and Tellurium-light.
This effect was observed in our PLD experiments where isolated islands or serpentine connected networks tended to occur at higher temperatures  (see Figure S1, Supporting Information), which is likely due to surface desorption.

Low temperature growth is attractive because the Te vapour pressure is lower, and its sticking coefficient is higher, which enables stoichiometric films to be grown. 
However, lower temperatures imply lower diffusivity of adatoms  which can hinder  intralayer and interlayer mass transport and lower the crystal quality.
This leads to amorphous or misoriented polycrystalline films with very small crystal grains, as shown in Figure~\ref{fig_6}(c).

Our experiments show that the PLD-grown films are much less sensitive to temperature than sputter-grown films. 
Moreover, others have shown that high quality PLD-grown \ch{Sb2Te3} vdW layered films are possible at extremely low temperatures, just 413~K, whereas sputtering produces polycrystalline films at this temperature, as shown in Figure~\ref{fig_6}(c) and (d)\citep{hilmi2017research, behera2018sb2te3}.
We hypothesise that this means that PLD adatoms have a higher diffusivity  at lower temperatures than sputtered adatoms.
Both PLD and sputtering are plasma-based deposition methods, but the plasma excitation energies are very different\citep{morfill1999plasma}. 
Moreover, the plasma intensity in PLD is orders of magnitude higher than in sputtering.
So the proportion of the ions in an excited state is much higher for PLD than sputtering. 
These excited state adatoms are likely to diffuse further than ground state atoms because the valence electrons, which are normally used to form bonds, are now lacking\citep{li2011role}. 
We propose that this explains why the crystal quality of PLD grown films is less sensitive to the substrate temperature and more sensitive to the seed layer.

It is also worth noting that neutral atoms separated by a distance $R$ attract each other via the weak $1/R^7$ vdW force. 
This is likely to be the case for sputtering where the majority of the atoms are not excited.
However,  in PLD where a larger proportion of the atoms are likely to be in an excited state, the force between atoms should scale as $1/R^4$, and this can be large enough to bind atoms into long range structures. 
In fact, provided atoms are in different excited states this relationship should still remain true\citep{kleppner1990apologies}.  
With this force scaling argument in mind, we should expect substantial differences in the structure resulting from sputtering and PLD.

\begin{figure}
\centering
\includegraphics[width=\textwidth]{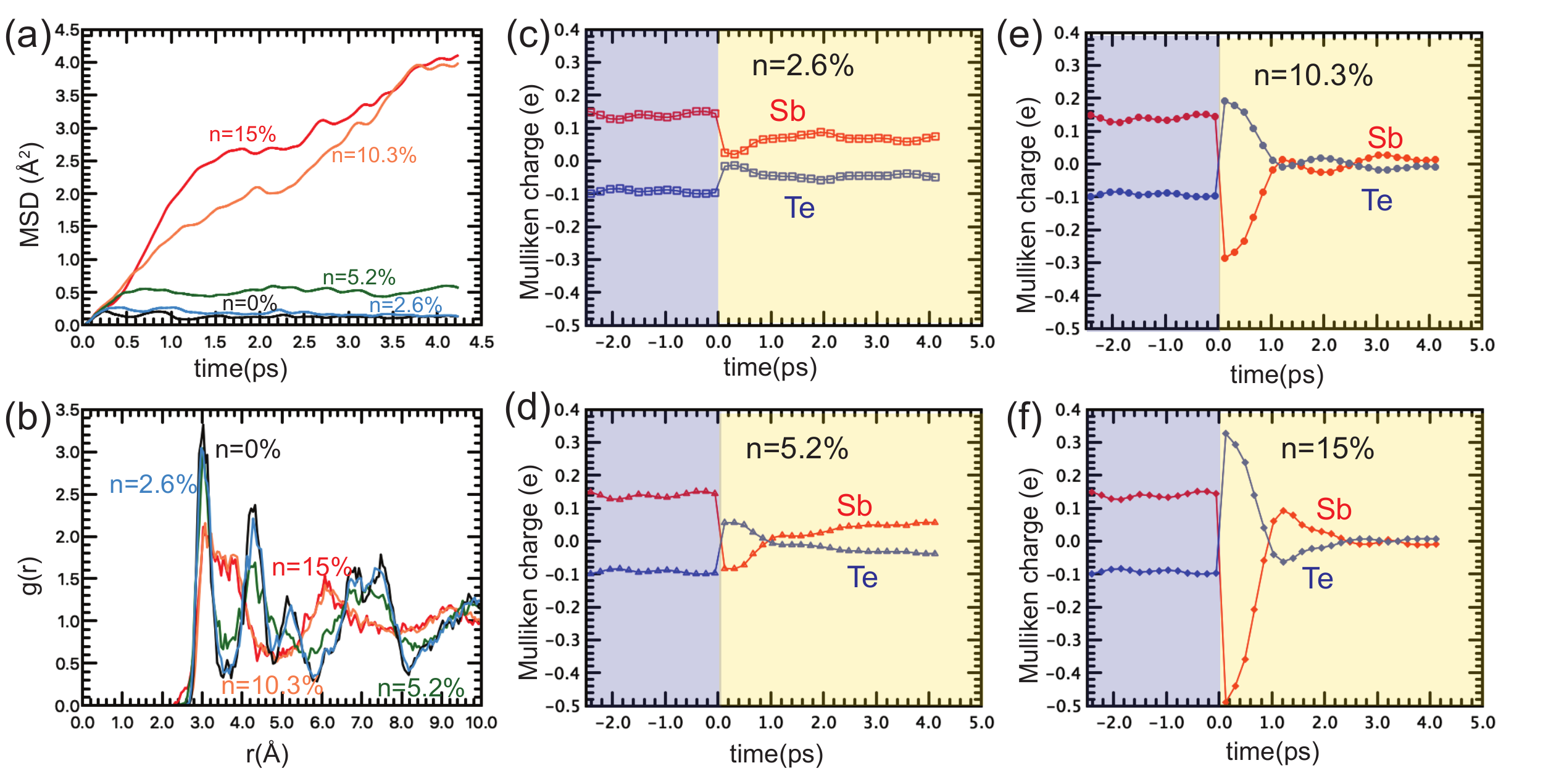}
\caption{Excitation effect on amorphisation at low temperature by NAQMD simulations. (a) Mean square displacement of Sb as a function of time. (b) Radial distribution function of \ch{Sb2Te3}. Cyan, green, orange and red curves correspond to excitation of n = 2.6\%, n = 5.2\%, n = 10.3\% and n = 15\% respectively at temperature 500 K. Black curve corresponds to the adiabatic MD at 500 K. n is percentage of excited electron out of the total electrons.
(c)--(f) Average Mulliken charge on Sb and Te as a function of time for 2 different excitation. Blue region corresponds to charges before excitation. Yellow region corresponds to the charges after excitation. Before excitation, Sb is positively charged and Te is negatively charged. Average charge on Sb and Te is shown by red and blue colour, respectively. }
\label{fig_7}
\end{figure}

To demonstrate the effect of the excited state on Sb and Te diffusivity during PLD growth, we performed non-equilibrium (excited state) DFT computations. 
Figure~\ref{fig_7}(a) shows the mean square displacement (MSD) for different proportions of valence electrons excited.  
We can see that the diffusivity of both Sb and Te atoms radically increases with the proportion of electrons excited.    
The resultant radial distribution function (RDF) shows significant  smearing when more than 10\% of the electrons have been excited. 
This is indicative of athermal melting at temperatures substantially below the equilibrium melting temperature. 
We now can affirm that PLD provides a means for atomic diffusivity to be high at the low temperatures where the Te sticking coefficient is also high; thus explaining how PLD can  be used to grow high quality and stoichiometric \ch{Sb2Te3} crystals at temperatures as low as 413~K \cite{hilmi2017research}. 
It also explains our observation that PLD is much less dependent on the growth temperature than sputtering -- i.e. the electronic excitation dictates the adatom diffusivity in PLD whereas the diffusivity is determined by the substrate temperature in sputtering. 
Moreover, these results suggest that excited state deposition methods are generally more likely to produce higher quality crystals when one or more of the adatoms has a low sticking coefficient at relatively low temperatures and hints that PLD is more suitable than MBE for growing similar Tellurium-based crystals. 

These excited state computations also allow us to follow the charge transfer between atoms as a function of time. 
A charge transfer process occurs for all the excitation levels studied (See Figure~\ref{fig_7}(c)--(f)).
Although the excited state lifetime is picosecond duration, long lived cascaded excitations can increase the effective lifetime of the excited state\citep{shang2018modeling}.
However, for weak excitation levels (n = 2.6\%)  Sb remains positively charged and Te remains negatively charged after the transfer.
In contrast, at high excitation levels, which we assume occurs in PLD, the charges are redistributed equally resulting in neutral Sb and Te atoms. 
This takes between 1 and 2~ps and depends on the proportion of electrons excited. 
We expect that these neutral atoms can move through a neutral \ch{Sb2Te3} network with minimal Coulomb interaction, resulting in a larger diffusivity and MSD, as seen in Figure~\ref{fig_7}(a) and (b). 
This enables adatoms to find low energy and stable positions required for high quality crystal growth and explains why  PLD-growth depends more strongly on the seed layer rather than the growth temperature.

DF/MD equilibrium simulations were also performed to study how the seed layer and ionisation influence the crystallisation of Sb and Te atoms.
The effect of the buffer was studied using a fixed monolayer of Te atoms, whilst the excited state atoms were modelled by removing the 10\% of the highest energy valence electrons from the band structure.
Ideally, we would like to apply NAQMD to study crystallisation.
However, for our model, the NAQMD crystallisation computation time is unfeasibly long. As a compromise, we decided to run long ground state DF/MD calculations with artificially excited electronic levels, and then reconfirm the general trends with shorter NAQMD simulations.

\begin{figure}
\centering
\includegraphics[width=\textwidth]{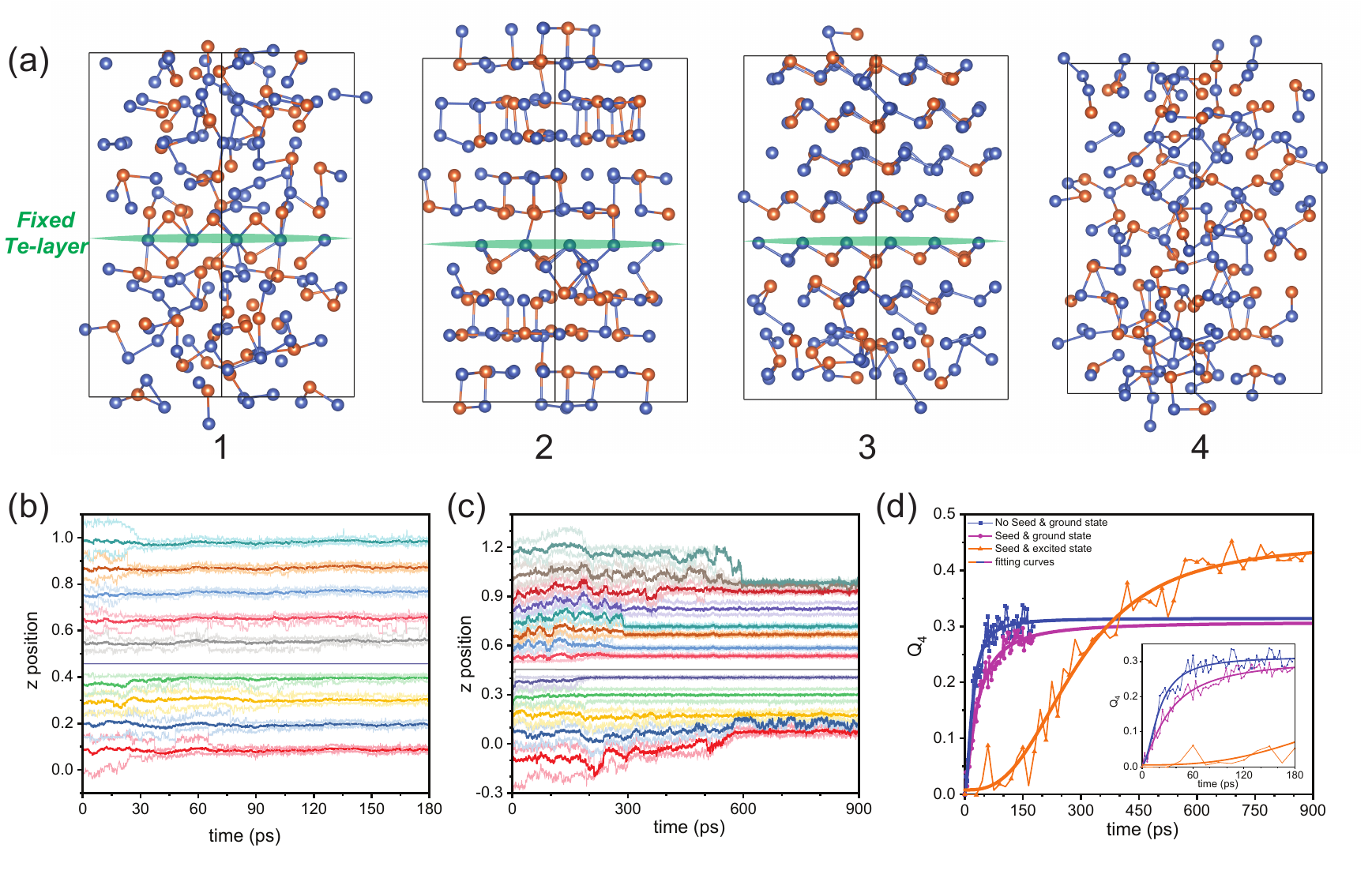}
\caption{Seed layer and excitation effect by DF/MD simulation. (a) crystallisation structure of initial and final state. Initial state after melting(a-1), final state of model with seed layer at ground state(a-2), with seed layer at excited state(a-3), without seed layer at excited state(a-4). Z-position evolution of simulation of structure with seed layer at ground state in 180 ps(b) and with seed layer at excited state in 900 ps(c). (d) Comparison of crystallinity by BOP. The inset figure shows a magnification of BOP curve in 180 ps. Blue square, purple circle and orange triangle correspond to model without seed layer at ground state, with seed layer at ground state and with seed layer at excited state, respectively.}
\label{fig_8}
\end{figure}

To model sputtered crystal film growth on a \ch{Sb2Te3} seed layer, where the majority of the atoms are not in an excited state, recrystallisation molecular dynamics simulations were run from a melt-quenched disordered structure (see Figure~\ref{fig_8}(a)-1) with a fixed layer of Te atoms (green area)  at a temperature of 800~K, as shown in Figure~\ref{fig_8}(b). 
Surprisingly, the recrystallised structure did not exhibit a layered structure after 180~ps, but recrystallised into a cubic layered crystal structure, as seen in Figure~\ref{fig_8}(a)-2. 

To model sputtered-growth on a substrate without a seed layer, a model was used with no fixed atoms. 
This amorphous structure also recrystallised into  a similar cubic layered structure (see Figure S3, Supporting Information). 
However, the recrystallisation time was slightly shorter and crystallinity marginally higher (see inset in Figure~\ref{fig_8}(d)). 
Due to the stochastic nature of crystallisation, this difference in crystallisation time between that with a Te layer and that without a fixed Te layer is unlikely to be significant. 

Importantly, when 10\% of the atoms were ionised and when a fixed Te atomic layer was included in the model, a highly crystalline and layered structure resulted.
These conditions model  that of PLD  with a \ch{Sb2Te3} seed layer, 
The structure gradually recrystallised  to form the ideal hexagonal \ch{Sb2Te3} layered structure.
This was possible despite mixed Sb and Te atomic layers, as shown in Figure~\ref{fig_8}(a)-3. 
Figure~\ref{fig_8}(c) shows the evolution of atomic positions along the c-axis of the hexagonal simulation cell (z-coordinate). 
The atoms closest to the seed layer tend to stabilise first; this is clear by the lower variation in the atomic positions.
After 600~ps most of atomic layers have formed and the variance in the atomic z-coordinate is reduced, thus indicating that the atoms have found a suitable low energy position in the crystal structure.
The BOP curves reveal that the excited state structure has a longer nucleation time and crystal growth only proceeds after 100~ps.
This is in contrast to the other curves where crystal growth proceeds immediately. 
This makes sense because ionisation depopulates valence band electrons, so it is more difficult for the atoms to form bonds, thus resulting in a higher diffusivity. 
So the atoms can wander around on the surface until they find a much lower energy position, which is much more likely to result in a more perfect crystal (see orange line in Figure~\ref{fig_8}(d)).
A similar effect is seen for the MSD in non-equilibrium DFT computations--see Figure \ref{fig_7}.
Note that the excited structure without the seed layer showing a disordered state did not crystallise during the 750 ps molecular dynamics simulation. Its final structure is shown in Figure~\ref{fig_8}(a)-4.
Generally, both the DF/MD with the 10\% highest energy electron removal, and the NAQMD computations qualitatively agree and indicate that electronic excitation increases the diffusivity of Sb and Te. However, we refrain from computing absolute quantities, such as adatom diffusivity, because the electronic excitation level is larger than what we might expect in reality.

It is this diffusivity increase that allows PLD to grow high quality \ch{Sb2Te3} vdW layered films at low temperatures, where the high Te sticking coefficient allows stoichiometric compositions
Figure \ref{fig_5} \& \ref{fig_8} show that the statistical analysis of the growth experiments, and the DF/MD models show that a  substantial improvement in the hexagonal crystal quality of \ch{Sb2Te3} can be achieved when both a seed layer and electronic excitation (PLD) is used to grow the films.
This is the case depicted in Figure \ref{fig_6}(d).
Thus there is a strong interaction between the excited state and the seed layer that influences the crystal quality;  a low quality crystalline  film results when there is no seed layer for PLD (excited state) grown films but high quality films are possible with no seeds for sputtering (less excited).
Moreover, these results provide a reasonable explanation for PLD being more suitable than sputtering for growing vdW layered crystals and their superlattices.

\section{Conclusion}

Previous growth studies on \ch{Sb2Te3}, as well as  other vdW layered materials, tended to generalise the findings of a single deposition method to other PVD methods. 
However, we have used sputtering, PLD, statistical design of experiments, and non-equilibrium DFT to unequivocally show that  the factors influencing the growth of simple binary chalcogenide crystals are very different for different growth methods, and therefore it is not possible to generalise the findings from one growth method to another.
We found that the quality of PLD grown films are insensitive to the growth temperature, and instead a seed layer significantly improves the crystal quality. 
We showed that this is due to the PLD plasma exciting \ch{Sb2Te3} valence electrons, which substantially increases the adatom diffusivity.
This higher diffusivity allows  crystal growth to occur at the low temperatures, which is necessary for Te atoms to stick to the substrate and achieve stoichiometric transfer.
The \ch{Sb2Te3} seed layer  provides a template from which the \ch{Sb2Te3} film can grow with a high degree of crystallographic orientation and this seems to be especially important for PLD-grown films.
This work highlights important differences between two plasma-based deposition methods that we suspect are applicable to not only tellurium-based chalcogenides but other narrow band gap semiconductors.

\section{Acknowledgment}
The authors thank Dr. V\'{a}clav Ocelik for the guidance of t-EBSD measurement. 
The SUTD research was funded by a Singapore MoE Project "Electric-field induced transitions in chalcogenide monolayers and superlattices", grant MoE 2017-T2-1-161. 
The USC research was supported as part of the Computational Materials Science Program funded by the U.S. Department of Energy, Office of Science, Basic Energy Sciences, under Award Number DE-SC0014607.
Ms Jing Ning is grateful for her MoE PhD scholarship.
The SUTD authors are grateful for space and facilities provided by the SUTD-MIT International Design Center (IDC).

\end{document}